\documentstyle[epsfig]{aipproc}
\newcommand{\nat}{Nature,~ }
\newcommand{\apjl}{Ap. J. Lett.,~}
\newcommand{\apj}{Ap. J., ~}
\newcommand{\physrep}{Physics Reports,~}
\newcommand{\aaps}{Astron. \& Astroph. Supp.,~}
\newcommand{\aap}{Astron. \& Astroph.,~}
\newcommand{\mnras}{Mon. Not. RAS, ~}

\begin{document}
\title{Gamma-Ray Bursts -  When Theory Meets Observations}

\author{Tsvi Piran}
\address{Racah institute for Physics, \\
The Hebrew University, Jerusalem Israel 91904}

\maketitle

\begin{abstract}

Gamma-Ray Bursts (GRBs) are the brightest objects observed. They
are also the most relativistic objects known so far. GRBs occur
when an ultrarelativisitic ejecta is slowed down by internal
shocks within the flow. Relativistic particles accelerated within
these shocks emit the observed gamma-rays by a combination of
synchrotron and inverse Compton emission. External shocks with
the circumstellar matter slow down further the ejecta and produce
the afterglow, which lasts for months. Comparison of the
predictions of this fireball model with observations confirm a
relativistic macroscopic motion with a Lorentz factor of $\Gamma
\ge 100$. Breaks in the light curves of the afterglow indicate
that GRBs are beamed with typical opening angles of a few
degrees. The temporal variability of the gamma-rays signal
provide us with the best indirect evidence on the nature of the
``internal engine'' that powers the GRBs and accelerates the
relativistic ejecta, suggesting accretion of a massive disk onto a
newborn black hole: GRBs are the birth cries of these black holes.
Two of the most promising models: Neutron Star Mergers and
Collapars lead naturally to this scenario.

--------------------------------------------------------------------------------

\end{abstract}

\section*{Introduction}

Gamma-Ray bursts - GRBs, short and intense bursts of
$\gamma$-rays arriving from random directions in the sky were
discovered accidentally more than thirty years ago. During the
last decade two detectors, BATSE on  CGRO and BeppoSAX have
revolutionized our understanding of GRBs. BATSE has demonstrated
\cite{BATSE92} that GRBs originate at cosmological distances in
the most energetic explosions in the Universe. BeppoSAX
discovered X-ray afterglow\cite{Costa97}. This enabled us to pinpoint the
positions of some bursts, locate optical \cite{vanP97} and
radio \cite{Frail97} afterglows, identify host galaxies and measure redshifts to
some bursts \cite{Metzger97}.

Since their discovery GRBs were among the prime topics of the
Texas Symposia. The high energy release and the rapid time scales
involved suggested immediately  association with relativistic
compact objects. The discoveries of BATSE and BeppoSAX confirmed
these expectations. These observations  have  established the
Fireball model demonstrating that GRBs are the most relativistic
objects known so far: GRBs involve macroscopic ultrarelativistic
flows with  Lorentz factors $\Gamma \ge 100$. Furthermore, while
the  ``central engines" that drive the relativistic flow and
power the GRBs are hidden we have excellent evidence that they
involve accretion onto a newborn black hole. GRBs are the birth
cries of these black hole.

I review, here,  the recent progress in our understanding of
GRBs, emphasizing, as appropriate for this conference, their
relativistic nature. I begin, in section \ref{texasny} with a
brief tribute to the 7th Texas symposium. This was the first
Texas meeting after the discovery of GRBs was announced and GRBs
were the highlight of the discussion there. I continue in section
\ref{fireball} with a brief exposition of the Fireball model (see
\cite{P99,P00} for details), confronting its predictions with the
observations. In \ref{engine} I summarize the implication of the
fireball model to the ``inner engines". Concluding remarks,
further predictions and open questions are discussed in section
\ref{conclusions}.

\section{The 7th (New York) Texas Symposium}
\label{texasny}

GRBs  were the hightlight of the Seventh (New York) Texas symposium
that took place in 1974. Five out of the 57 talks   were
devoted to GRBs (this record was repeated only
in this symposium with 3 out of 29 talks): Two observational reviews, a theoretical
review, a theoretical model and even a description of an automated system for searching
for optical transients accompanying GRBs!

M. Ruderman \cite{Ruderman75} reviewed  the theory\footnote{This
review enumerates more than thirty models proposed during the
short time passed since the announcement of the discovery. It is
remarkable (Ruderman, 1998, private communication) that today we
know that none of these models is even remotely relevant.}.
emphasizing the {\it compactness problem}: If GRBs are
cosmological then the energy budget and the time scales seem to
be incompatible with the observed non thermal spectrum of the
bursts. The argument is simple: the variability time scale,
$\delta t$, imposes an  upper limit on the size ($R \le c \delta
t$). The observed flux and the assumed (cosmological) distance
determine the energy. Together these yield an extremely large
lower limits on the photons  density within the source and on the
optical depth for pair creation by the energetic photons.
Pairs would be copiously produced and the source
would be optically thick. The observed optically thin
spectrum is impossible. Ruderman points out, however, that
relativistic effects would change this conclusion. If the source
is moving relativistically the relations between the time scale
and the implied distance are modified (by a factor of
$\Gamma^{2}$). Furthermore, photons that are observed with energy
below 500$\Gamma$keV have  energy below 500keV in the source rest frame
and could not produce pairs. These ideas
lay the foundation for the current Fireball model. Recent
observations have indeed confirmed ultrarelativistic motion in
GRBs, showing at least in one case $\Gamma \ge 100$.

\section{The Fireball Model, Predictions and Confirmations}
\label{fireball}
One can never prove a scientific theory. However we gain
confidence in a theory when its predictions are confirmed by
observations.
I discuss, here, the predictions of the Fireball
model (specifically of the Fireball-Internal-External shocks
model) and their confirmation by numerous observations. My goal
is to demonstrate the success of this model. While some of the
specific
observations could certainly be interpreted  within
other theories I strongly believe that the bulk of those
observations tell us that this is the correct model.

\begin{figure}[b!]  
\centerline{\epsfig{file=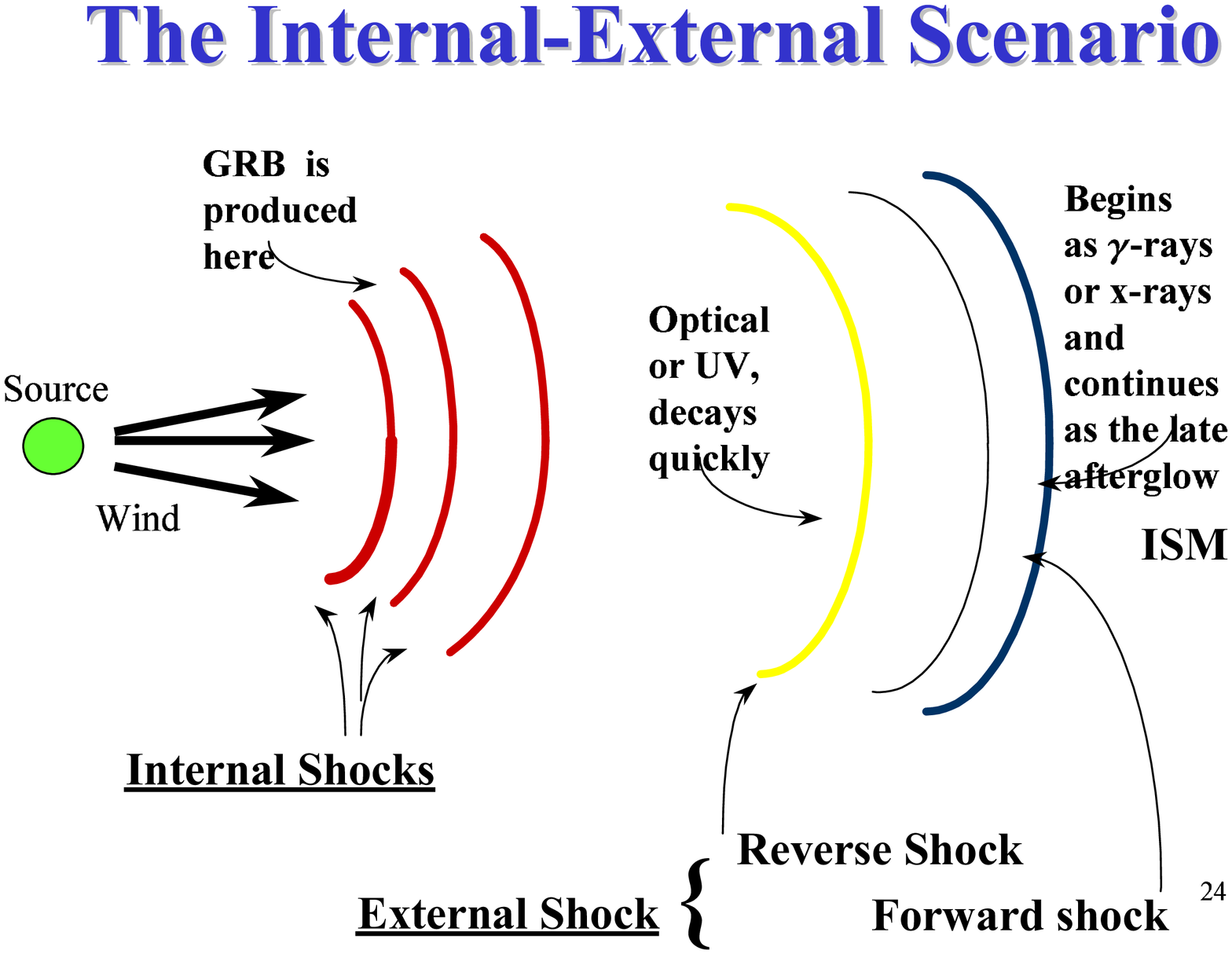,height=3.5in,width=4.5in}}
\vspace{10pt}
\label{schematic}
\end{figure}

The Fireball model asserts that GRBs are produced when the
kinetic energy (or Poynting flux) of a relativistic flow is
dissipated by shocks\footnote{Given the low densities involved
these shocks, like SNR shocks,  must be collisionless.}. These shocks accelerate electrons
and generate strong magnetic fields. The relativistic electrons emit
the observed $\gamma$-rays via synchrotron or SSC.
There are two variants of this model: The External Shocks model
\cite{MR93a} assumes that the shocks are between the relativistic flow and the
surrounding circumstellar matter. The Internal Shocks model
\cite{NPP92,RM94}
assumes that the flow is irregular and the shocks take place
between faster and slower shells within the flow.
According to the Internal-External shocks model \cite{SP97} both
kinds of shocks take
place: Internal shocks are responsible for the GRB
while external shocks produce the longer lasting afterglow
(see Fig. 1) . Both
shocks occur at relatively large distances ($10^{13}-10^{14}$cm
for internal shocks and $10^{14}-10^{16}$cm for external shocks)
from the source that generates the relativistic flow. The
observed radiation from the GRB or from the afterglow
reflects only the conditions within these shocks. We have only
indirect information on the nature of the ``inner engines".

Fig. 1 depicts a schematic picture of the
Internal-External shocks model. An inner engine produces an
irregular wind. The wind varies on a scale $\delta t$ and its
overall duration is $T$. The variability scale $\delta t$
corresponds to the variability
scale observed in the GRB light curve \cite{KPS97}
thus, $\delta t \sim 1$ sec. Internal shocks take place at
$R \approx \delta t \Gamma^2  \approx 3 \cdot 10^{14} {\rm cm}
~(\delta t/1~ sec) (\Gamma/100)^2$.  External shocks shocks become significant
and the blast wave
that propagates into the circunstellar matter and produces the afterglow begins
at $\sim 10^{16}$cm (see \cite{SaP95,P99} for details).
At this initial stage there is also a short lived reverse shock that propagates
into the ejecta. This reverse shock is responsible to the
prompt optical emission observed in GRB990123.

I review  now the predictions of this model
and confront them with observations.

{\bf  $\bullet$ Relativistic Motion - Predictions:}
Relativistic motion is the key ingredient of the Fireball model.
Relativistic motion  arises naturally (a Fireball forms) when a large amount of energy is produced in a
compact region with $E/Mc^2 \gg 1$ \cite{Goodman86,Pac86,SP90}.
This relativistic motion overcome the
compactness problem. Various estimates \cite{FenimoreEH93,WoodsL95,P95,SaLi00},
of the Lorentz factor $\Gamma$ based on the compactness problem
lead to comparable values, $\Gamma \sim 100$, (see \cite{SaLi00} for a
critical review).

{\bf  $\star$ Relativistic Motion - Observations I:}
The radio afterglow observations of GRB 970508 provided the first verification
of relativistic motion. The radio light curve (in 4.86Ghz) variesd strongly
during the first month. These variations died out later.
Even before this transition
Goodman \cite{Goodman97} interpreted these variations as
scintillations. The observation of a transition after one month enabled
Frail et al., \cite{Frail97} to estimate the size of the afterglow
at this stage as $\sim 10^{17}$cm. It immediately follows that the afterglow expanded
relativistically.
Additionally, the source is expected to be optically thick in radio
\cite{KP97} leading to a $\nu^2$ rising spectrum at these frequencies.
The observed flux from the source
enables us (using the black body law) to estimate the size of the
source. As predicted the radio spectrum increases like $\nu^2$.
The size estimated with this method agrees
\cite{Frail97} with the one derived by the scintillations estimate
implying as well a relativistic motion.

{\bf  $\bullet$ The Afterglow - Predictions:} The Afterglow -
lower frequency emission that follows GRBs was
one of the earlier predictions of the Fireball model
\cite{PaczynskiR93,MR93,MR97,Vietri97}.
Paczynski and Rhoads \cite{PaczynskiR93} predicted radio afterglow on
the basis of  the analogy between
external shocks and SNRs. Later Meszaros and Rees \cite{MR93,MR97}
performed detailed calculations of  multi-wavelength afterglow.
Vietri \cite{Vietri97} predicted soft x-ray  afterglow as a test for
the external shocks model. These predictions were done in the
context of the external shocks
model. According to this model the GRB arises due to a shock
between a relativistic flow and a circumstellar medium. In this
case the emission observed at time $t_{obs}$ arises when $t_{obs}
\approx R/2 \Gamma^2$. Later emission is related to lower
$\Gamma$ (and larger emission radii) and hence to lower observed
frequencies.  This afterglow would be a
direct extrapolation of the GRB and its basic features should be
strongly correlated with the properties of the corresponding GRB.

The theory of the afterglow is well understood. Blandford and
McKee \cite{BM76} have worked out (already in the seventies!) the theory of an adiabatic
relativistic blast wave. They show that (as long as the flow is
ultrarelativistic, $\Gamma \gg 1$), the blast wave is self
similar, the relativistic analog of the well known Sedov-Taylor
solution. Electrons are accelerated to relativistic velocities by
the shocks and their interaction with the magnetic field leads to
synchrotron radiation. This provides an excellent model for the
observed emission \cite{SPN98}. Overall we have a simple theory
characterized by five parameters: the total energy, $E_0$, the
ambient density, $n_0$, the ratio of the electrons and magnetic
fields energy density to the total energy density, $\epsilon_e$,
$\epsilon_B$ and the exponent  of the electrons' energy
distribution function $p$. An additional sixth parameters, the
exponent of the circumstellar density distribution, $n$, arises
in cases when the external matter density ($\rho \propto r^{-n}$).
Most notable is $n=2$ corresponding to a  pre-GRB wind expected
in some models \cite{Chevalier00}. This rather simple theory
predicts a robust relations between $\alpha$ and $\beta$ the
exponents describing the flux as a function of  frequency, $F_\nu
\propto t^{-\alpha} \nu^{-\beta}$. At the high frequencies,
above the cooling frequency, we have (for $n=0$), $\alpha =
(3p-2)/4$ and $\beta=p/2$.

{\bf  $\star$ The Afterglow - Observations:} On Feb 28 1997, in a
wonderful anniversary celebration for SN87A, BeppoSAX detected
x-ray afterglow from GRB 970228. The exact position given by
BeppoSAX led to the discovery of optical afterglow \cite{vanP97}.
Radio afterglow was detected in GRB 970508 \cite{Frail97}. By now
more than thirty  x-ray afgterglows have been observed. About
half of these have optical and radio afterglow as well and in
most of those the host galaxy has been discovered.

Most x-ray and optical afterglow decay as power laws with
$\alpha \sim 1.2$ and  $\beta \sim 1.2$, in
excellent agreement with the predictions of the simplest afterglow
model: An adiabatic  Blandford-McKee hydrodynamics with
Synchrotron emission \cite{SPN98}. As $\alpha$ and $\beta$ are
determined by $p$ the electron's distribution power law index,
these observations suggest that as predicted \cite{SPN98,SNP96},
$p\approx 2.5$ and it  is fairly invariant from one burst to
another \cite{PanaitescuK01}. A simultaneous spectral fit for
GRB980508, all the way from the radio to the x-ray also agrees
with this picture \cite{WijersGalama99}.

{\bf $\bullet$ The GRB-Afterglow Transition - Predictions:}
The rapid time variability seen in most GRBs cannot be produced by
external shocks \cite{SP97,Fenimore96}. This leaves internal shocks as
the only viable model!
Shortly before the discovery of the afterglow from GRB970228,  Sari and Piran
\cite{SP97} pointed out that
afterglow should arise also within the
internal shocks scenario.  The efficiency of
the internal shocks depends on the parameters of the flow, most
specifically on the variability of the Lorentz factor between
different shells \cite{KPS97,Daigne98}. Even in the
most efficient cases a significant fraction of the energy remain
as kinetic energy. Sari and Piran \cite{SP97} suggested that
this energy would be dissipated later by
interaction with the surrounding matter and produce an afterglow.  Within this
Internal-External shocks model the GRB is produce by internal
shocks while the afterglow is produced by external shocks. The
predictions of this model for the afterglow are similar to those
of the External shocks model. However, a critical difference is
that here the  afterglow is not an extrapolation of the GRB.

The internal shocks take place at a distance $R_{IS} \sim \delta t
\Gamma^2 \sim 10^{14}$cm. These shocks last as long as the inner
engine is active. The
typical observed time scale for this activity  $\sim 50 $sec (for
long bursts) and $\sim 0.5$sec (for short ones). External shocks
begin at $R_{Ex} \sim 10^{16}$cm. If $R_{Ex} /\Gamma^2 \le T$ this happens
while internal shocks are still going on and the afterglow
overlaps the late part of the GRB. At the early time the
afterglow emission peaks in the high x-rays contributing also to
the observed $\gamma$-ray flux. We expect, therefore,  a
transition within the GRB from hard (pure GRB) to softer and
smoother (GRB and afterglow) signal.

{\bf $\star$  The GRB - Afterglow Transition - Observations:}
The extrapolation of the x-ray afterglow fluxes backwards generally
does not fit the $\gamma$-ray fluxes. Moreover there is no direct
correlation between the $\gamma$-ray fluxes and the x-ray (or optical)
afterglow fluxes. This result is in a nice agreement with the
predictions of the Internal - External shocks scenario in which
the two phenomena are produced by different effects and one
should not expect a simple extrapolation to work.

The expected GRB afterglow transition have been observed in
several cases. The first observation took place (but was not
reported until much latter) already in 1992 \cite{Burenin99}.
Recent BeppoSAX data shows a rather sharp transition in the
hardness that takes place several dozen seconds after the
beginning of the bursts \cite{Costa99}. This transition is
seen clearly in  the different energy bands light curves of GRB990123
and in GRB980923 \cite{Giblin99}.
Connaughton \cite{Connaughton98} have averaged the light curves of
many GRBs and discovered  long and soft tails: the early x-ray
afterglow.

{\bf  $\bullet$ The Prompt Optical Flash -  Predictions:} The collision
between the ejecta and  the surrounding medium produces two
shocks. The outer forward shock propagates into the ISM. This
shock develops later into the self similar Blandford-McKee blast
wave that drives the afterglow. A second shock, the reverse
shock,  propagates into the flow. This reverse shock is short
lived. It dies out when it runs out of matter as it reaches the
inner edge of the flow. While it is active, it is a powerful
source of energy. Comparable amounts of energy are dissipated by
the forward and by the reverse shocks \cite{SaP95}. We expect
that the system is radiative at this stage, namely most of the
energy converted by the shock is radiated away.

Sari and Piran  predicted at the First Rome Meeting
(Oct 1998) an intense (brighter than 11th magnitude) prompt
optical flash from this reverse shock \cite{SP99c}. Previous work \cite{MR97}
done prior to the discovery of the afterglow  considered various
possibilities and estimated the magnitude of the prompt optical
flash to be anywhere from 9th to 19th magnitude. The
observations of the afterglow constrained severely the relevant
models and the relevant parameter space.  With the new data the
constrained model led to a clear
prediction with a narrow range of magnitude.  At that time this
prediction was almost conflicting with upper limits given by
systems like LOTIS  and ROTSE.

One prediction that  was, unfortunately, missed:
prompt  radio emission from the reverse shock. This radio
emission should be short lived, like the burst and should have
initially an optically thick component that becomes optically
thin later.

{\bf $\star$ The Prompt Optical Flash - Observations:}
In Jan 23 1999 just three month after this prediction ROTSE
recorded six snapshots of optical emission from
GRB990123\cite{ROTSE} .  Three of those were taken while the
burst was still emitting $\gamma$-rays. The other three snapshots
spanned a couple of minutes after the burst. The second snapshot,
taken 70sec after the onset of the burst corresponds to a 9th
magnitude signal.  A comparison of these optical observations
with the $\gamma$-rays and x-rays light curves (see e.g.
\cite{P00})  shows that the optical emission does not correlate
with the $\gamma$-rays pulses. The optical photons and the
$\gamma$ rays are not emitted by the same photons
\cite{SP99b,ROTSE}. The optical pulses peak some 70sec after the
onset of the burst simultaneously with a late peak in the soft
x-ray emission.

Radio observations of GRB990123 revealed a short lived radio
pulse. This emission can be explained as coming from the reverse
shock \cite{SP99b}.  Using the parameters of the reverse shock
derived from the optical flash Sari and Piran \cite{SP99b}
estimated the magnitude of this radio emission. The theoretical
curve and the observations are in excellent agreement (see e.g.
\cite{P00}). Note that the theoretical curve was calculated just
from the optical flash data and it was not  ``fitted" in any way
to the observed data. While prompt optical flashes were not
detected in other bursts, short lived radio flashes have been
detected in GRB000926 and in GRB970828.

{\bf $\star$ Relativistic Motion II - Observations:}
The radio emission from  GRB970508 showed relativistic motion in
its afterglow. However, the significant observations were done
one month after the burst and at that time the motion was only
``mildly" relativistic with a Lorentz factor of order a few. The
observations of GRB990123 enabled us to obtain three independent
estimates of the ultrarelativistic Lorentz factor at the time that the ejecta hits
first the ISM \cite{SP99b}. First the time delay between the GRB
and the optical flash suggests $\Gamma \sim 200$. The ratio
between the emission of the forwards shock (x-rays) and the
reverse shock (optical) gives another estimate of $\Gamma \sim
70$. Finally the fact that the maximal synchroton frequency of the
reverse shock was below the optical band led to $\Gamma \sim 200$.
The agreement between these three crude and independent estimates is reassuring.

These observations provide us also with an estimate of the
position of the external shocks, $\sim 10^{15}$cm at 70 seconds
after the bursts. It is an impressive measurement considering the
fact that the distance to this burst is $\sim 3.5 Gpc$. The
corresponding angular resolution is $10^{-13}$ or $\sim 50
{\rm nanoarcsec}$.

{\bf  $\bullet$ Jets - Predictions:} With redshift measurements it became
possible to obtain exact estimate the total energies involved.
While the first burst GRB970508 required a modest value of $\sim
10^{51}$ergs, the energies required by other bursts were alarming,
$3 \times 10^{53}$ergs for GRB981226 and $4 \times 10^{54}$ergs
for GRB990123, and unreasonable for any simple compact object
model. These values suggested that the  assumed isotropic
emission was wrong and GRBs are beamed. Significant beaming would
of course reduce, correspondingly the energy budget.

Beaming was suggested even earlier as it arose naturally in some
specific models. For example the binary neutron star merger has a
natural funnel along its rotation axis and one could expect that
any flow would be emitted preferably along this axis. The
Collapsar model also requires beaming, as only a concentrated
beamed energy could drill a whole through the stellar envelope
that exists in this model.

Consider a relativistic flow with an opening angle $\theta$. As
long as $\theta > \Gamma^{-1}$ the forwards moving matter doesn't
``notice" the angular structure  and the hydrodynamics is ``locally" spherical
\cite{P94}. The radiation from each point is beamed into a
cone with an opening angle $\Gamma^{-1}$.  It is impossible to
distinguish at this stage a jet from a spherical expanding shell.
When $\theta \sim \Gamma^{-1}$ the radiation starts to be beamed
sideways. At the same time the hydrodynamic behaviour changes and
the material starts expanding sideways. Both effects lead to a
faster decrease in the observed flux,  changing $\alpha$, the
exponent of the decay rate of the flux to: $\alpha=p/2$.
Thus we expect a break in
the light curve and a new relation between $\alpha$ and $\beta$
after the break \cite{Rhoads99,SPH99,PM99}. The magnitude of the
break and the duration of the transition will change if the jet is
expanding into a wind with $r^{-2}$ density profile \cite{PanaitescuK00}.
The break is expected to take place at $t_{jet} \approx
6.2 (E_{52}/n_0)^{1/3} (\theta/0.1)^{8/3}$hr \cite{SPH99}.
Recently numerical simulation \cite{Granot01} have
shown that the break appears in a more realistic calculations,
even though the numerical results suggest that the analytical
model developed so far are probably too simple.

{\bf  $\star$ Jet - Observations:} GRB980519 had unusual values for
$\alpha=2.05$ and $\beta=1.15$. These values do not fit the
"standard" spherical afterglow model\footnote{A possible
alternative fit is to a wind (n=2) model but with a unsual high
value of $p=3.5$}. However, these values are in excellent
agreement with a sideway expanding jet \cite{SPH99}. The simplest
interpretation of this data is that we observe a jet during its
sideway expansion phase (with $p=2.5$). The jet break
transition from the spherical like phase to this phase took place
shortly after the GRB and it was not caught in time.
The light curves of GRB990123 shows, however,
a break at $t\approx 2$days \cite{Kulkarni99}. This break is interpreted as a jet
break, corresponding to an opening angle $\theta \sim 5^o$.
Another clear break was seen in GRB990510
\cite{Harrison99,Stanek99}.

The brightest bursts, GRB990123 and GRB980519 gave the first
indications for jet like behaviour \cite{SPH99}. This suggested
that their apparent high energy was due to the narrow beaming
angles. A compilation of more bursts with jet
breaks suggests that all bursts have a comparable energy $\sim
10^{51}$ergs and the variation in the observed energy is mostly
due to the  variation in the opening angles $\theta$
\cite{Frail01,Kulkarni01,PanaitescuK01}

\section{The Inner Engines}
\label{engine}

The Fireball model tells us how GRBs operate. However, it does
not answer the most interesting astrophysical question: what
produces them? which astrophysical process generates the
energetic ultrarelativistic flows needed for the Fireball model?
Several observational clues  help us  answer these
questions.
 \\ \indent {\bf $\bullet$ Energy:} The total energy involved is large $\sim 10^{51}$ergs,
  a significant fraction of the binding energy of a
  stellar compact object. {\it The ``inner engine" must be able to
  generate this  energy and
  accelerate  $\sim 10^{-5}M_\odot$ to relativistic velocities.}
   \\ \indent {\bf $\bullet$ Beaming:} Most GRBs are
  beamed with typical opening angles $0.02<\theta<0.2$.
   {\it The ``inner engine" must be
  able to collimate the relativistic flow.}
   \\ \indent {\bf $\bullet$ Long and Short Bursts:} The bursts are divided to two
  groups according to their overall duration. Long bursts with $T>2$sec
  and short ones with $T<2$sec.
   \\ \indent{\bf $\bullet$ Rates:} GRBs take place once per $10^7 (4/\theta^2)$yr per
  galaxy. {\it GRBs are very rare at about 1/1000 the rate of
  supernovae.}

\noindent The Fireball Internal-External shocks model provides us with
another key clue:
  \\ \indent{\bf $\bullet$ Time Scales:} The variability time scale,
  $\delta t$, is at times as short
  as 1ms. The overall duration, $T$, is of the order of 50sec.
According to the internal shocks model these time scales are
determined by the ``inner engine". {\it $\delta t \sim$ msec
suggests a compact object. $T \sim 50$sec  is much longer than the
dynamical time scale, suggesting a prolonged
activity.\footnote{The ratio $\delta t/T \ll 1$ for short bursts
as well \cite{NakarP01}}. This rules out any ``explosive" model
that release the energy in a single explosion.}

\noindent
The internal shocks model requires two (or possibly three
\cite{Ramirez00,NakarP01a}) different
time scales operating within the ``inner engine".
These clues, most specifically the last one suggest that GRBs
arise due to accretion of a massive ($\sim 0.1 m_\odot$) disk
onto a compact object, most likely a newborn black hole. A
compact  object is essential because of
the short time scales.
Accretion is needed to produce the two different time scales, and
in particular the prolonged activity. A massive ($\sim 0.1 m_\odot$)
disk is required
because of the energetics. We expect that such a  massive disk can
form only simultaneously with the formation of the compact object.
This leads to the conclusions that GRBs accompany the formation of
black holes. This model is supported by the observations of relativistic
(but not as relativistic as in GRBs) jets in AGNs, which
are powered by accretion onto black holes.
This system is capable of
generating  collimated relativistic flows even though we don't understand how.

An important alternative to accretion is Usov's
model \cite{Usov92} in which the relativistic flow is mostly
Poynting flux and it  is driven  by the
magnetic and rotational energies  of a newborn rapidly
rotating neutron star. However  this model
seems to fall short by an order of magnitude of the energy required.

Several scenarios could lead to a black hole - massive accretion
disk system. This could include mergers (NS-NS binaries
\cite{Eichler_LPS89,NPP92}, NS-BH  binaries
\cite{Pac91}  WD-BH binaries \cite{Fryer_WHD99}, BH-He-star
binaries \cite{Fryer_Woosley98}) and models
based on ``failed supernovae'' or ``Collapsars''
\cite{Woosley93,Pac98,MacFadyen_W99}. Narayan et al.
\cite{NarayanPK01} have recently shown that accretion theory
suggests that from all the above scenarios only Collapsars could
produce long bursts and only NS-NS (or NS-BH) mergers could produce short
bursts.


Additional  indications  arise from afterglow observations. One
has to use these clues with care.  Not all GRBs have afterglow
(for example,  so far afterglow was not detected from any
short burst) and  it is not clear whether these clues are
relevant to the whole GRB populations. These clues seem
to suggest a GRB-SN connection:
   \\ \indent{\bf $\bullet$ SN association:}
   Possible association of GRB980425 with SN98bw \cite{GalamaSN98}
  and  possible SN signatures in the afterglows of
  GRB970228 \cite{Reichart99} and GRB980326 \cite{Bloom99}.
   \\ \indent{\bf $\bullet$ Iron lines:} have been observed
  in some x-ray afterglows \cite{Piro00}. Any model explaining them requires a
  significant amounts of iron at rest near those GRBs.
     \\ \indent {\bf $\bullet$ Association with Star formation:}  GRBs seem to follow the star formation rate.
  GRB are located within  star forming
  regions in star forming Galaxies \cite{Pac98,Kulkarni01}.
     \\ \indent {\bf $\bullet$ GRB distribution:}
     GRBs are distributed within galaxies. There is no evidence for
  GRBs kicked out of their host galaxies  \cite{Bloom01,Kulkarni01}
  as would be expected for NS-NS mergers  \cite{NPP92}.

All these clues  point out towards a SN/GRB association and
towards the Collapsar model. However,
the situation is not clear cut. The association of GRB980425 with
SN98bw is uncertain.  There are alternative explanations to the
bumps in the afterglows of GRB970228 and GRB980326 \cite{Esin00}.
Iron is produces in Supernovae. But there is no simple
explanation what is iron at rest doing around the GRB (see
however, \cite{Vietri01}). The association with star formation and the
distribution of GRBs within galaxies is real but all that it
indicates is short lived progenitors. One cannot rule out
a short lived binary NS  population \cite{TutukovY94}
which would mimic this behaviour. Even worse, there are some indication that
seem incompatible with the SN association:
     \\ \indent {\bf $\bullet$ No Windy Afterglow:} No evidence for a wind (n=2)
  in any of the afterglow light curves? Furthermore,
  most fits for the afterglow parameters show low
  ambient density \cite{PanaitescuK01,Kulkarni01}.
     \\ \indent{\bf $\bullet$ No Jets:} Some GRBs don't show evidence for a jet or have very
  wide opening angles \cite{PanaitescuK01,Kulkarni01}, this would be incompatible
  with the Collapsar model.

\section{Conclusions, Predictions and Open Questions}
\label{conclusions}

There is  an ample observational  support For the Fireball model.
It also has several other predictions. The most interesting ones
are those concerning
the very early afterglow and the GRB-afterglow transition.
The early afterglow phase is radiative and a detailed
look at the first hour  of the afterglow should show the
radiative to adiabatic transition. It should also show (mostly in radio)
small bumps  corresponding to refreshed shocks \cite{KP00a} which would
enable us to learn more on the nature of the flow produced by the
``inner engine". With an operational HETE II and Swift in the not
too distant future we hope that this crucial phase  will be explored soon.
Another  prediction of  a ring  structure of the afterglow\cite{GPS99c}
will have to wait, however,  to futuristic ultrahigh resolution
detectors.

We know how GRBs are produced. We are less certain what produces
them. We can trace backwards the evolution at the source from the
observations of the emitting regions to an accretion disk - black  hole system.
The traces from this point backwards are less clear.
Theoretical considerations \cite{NarayanPK01} suggest that only
Collapsars can produce the disk-black hole systems needed for
long bursts while only NS-NS (or possibly NS-BH) mergers can
produce the systems needed for short bursts. These conclusions
are supported by the afterglows  observations
that suggest SN/GRB association for the long burst population.
However, the picture is far from clear yet. While the information
on the location of the bursts points out towards the SN
connection the  physical conditions within the
afterglow indicate a low circumstellar density   and does not show
any indication for the almost inevitable pre explosion wind.
The origin of the iron lines
is still mysterious and confusing.

The Fireball model  has still many open
questions. Some are concerned  with the physics of the fireball model:
How do the collisionless shocks work?
How are the electrons accelerated and how are the magnetic fields amplified?
How do jets expand sideways?
What controls the ``typical" emission to be in the soft $\gamma$-ray
region? Other questions deal with more astrophysical issues like:
What happens in all the cases (like WD-BH merger) in which
a GRB almost form but the conditions are not exactly right?
What distinguishes between the prgenitor of a
GRB-``failed supernova"  and the progenitor of a successful supernova with no GRB?
Finally,  we have, of course, the
sixty four thousand dollars question: How does
the ``inner engine" accelerates  the ejecta to relativistic velocities?.


\end{document}